\documentclass[aps,prb,twocolumn,superscriptaddress,groupedaddress,showpacs]{revtex4}

\usepackage{graphicx}  
\usepackage{amssymb}   
\usepackage{amsmath}
\usepackage{color}
\usepackage{times}
\usepackage{bm}
\usepackage{bbm}

\newcommand{\AR}[1]{#1}

\newcommand{\br}{\mathbf{r}}

\newcommand{\be}{\mathbf{e}}

\newcommand{\bT}{\mathbf{T}}

\newcommand{\bM}{\mathbf{M}}
\newcommand{\bR}{\mathbf{R}}
\newcommand{\bB}{\mathbf{B}}
\newcommand{\bG}{\mathbf{G}}
\newcommand{\bb}{\mathbf{b}}
\newcommand{\bF}{\mathbf{F}}

\newcommand{\bV}{\mathbf{V}}

\newcommand{\uu}[1]{{{\mathbf{#1}}}}
\newcommand{\complex}{i}
\newcommand\id{\ensuremath{\mathbbm{1}}}

\begin{document}
\title{Inertia, diffusion and dynamics of a driven skyrmion}

\author{Christoph Sch\"utte}
\affiliation{Institut f\"ur Theoretische Physik, Universit\"at
zu K\"oln, D-50937 Cologne, Germany}

\author{Junichi Iwasaki}
\affiliation{Department of Applied Physics, University of Tokyo, 7-3-1, Hongo,
Bunkyo-ku, Tokyo 113-8656, Japan}

\author{Achim Rosch}
\affiliation{Institut f\"ur Theoretische Physik, Universit\"at
zu K\"oln, D-50937 Cologne, Germany}

\author{Naoto Nagaosa}
\email{nagaosa@riken.jp}
\affiliation{Department of Applied Physics, University of Tokyo, 7-3-1, Hongo,
Bunkyo-ku, Tokyo 113-8656, Japan}
\affiliation{RIKEN Center for Emergent Matter Science (CEMS), 
Wako, Saitama 351-0198, Japan}

\date{\today}


\begin{abstract}
Skyrmions recently discovered in chiral magnets are a promising candidate for magnetic storage devices because of their topological stability, small size ($\sim 3-100$nm), and ultra-low threshold current density ($\sim 10^{6}$A/m$^2$) to drive their motion.  However, the time-dependent dynamics has hitherto been largely unexplored. Here we show, by combining the numerical solution of the Landau-Lifshitz-Gilbert equation and the analysis of a generalized Thiele's equation, that inertial effects are almost completely absent in skyrmion dynamics driven by a time-dependent current. In contrast, the response to time-dependent magnetic forces and thermal fluctuations depends strongly on frequency and is described by a large effective mass and a (anti-) damping depending on the acceleration of the skyrmion. Thermal diffusion is strongly suppressed by the cyclotron motion and is proportional to the Gilbert damping coefficient $\alpha$. This indicates that the skyrmion position is stable, and its motion responds to the  time-dependent current without delay or retardation even if it is fast. These findings demonstrate the advantages of skyrmions as information carriers. 
\end{abstract}

\maketitle

\section{Introduction} 
\label{sec:Introduction}
Mass is a fundamental quantity of a particle determining its mechanical inertia and therefore the speed of response to external forces. Furthermore, it controls the strength of quantum and thermal fluctuations. For a fast response one usually needs small masses and small friction coefficients which in turn lead to large fluctuations and a rapid diffusion. Therefore, usually small fluctuations and a quick reaction to external forces are not concomitant. However a ``particle'' is not a trivial object in modern physics, it can be a complex of energy and momentum, embedded in a fluctuating environment. Therefore, its dynamics can be different from that of a Newtonian particle. This is the case in magnets, where such a ``particle'' can be formed by a magnetic texture \cite{Hubert98magnetic, Malozemoff79magnetic}. A skyrmion \cite{Skyrme61nonlinear, Skyrme62unified} is a representative example: a swirling spin texture characterized by a topological index counting the number of times a sphere is wrapped in spin space. This topological index remains unchanged provided spin configurations vary slowly, i.e., discontinuous spin configurations are forbidden on an atomic scale due to high energy costs. Therefore, the skyrmion is topologically protected and has a long lifetime, in sharp contrast to e.g. spin wave excitations which can rapidly decay. Skyrmions have attracted recent intensive interest because of their nano-metric size and high mobility \cite{Bogdanov89thermo, Muhlbauer09skyrmion, Yu10real, Heinze11spontaneous, Seki12observation, Fert13skyrmion, Lin13driven, Lin13particle, Iwasaki13constricted, Iwasaki2014colossal}. Especially, the current densities  needed to drive their motion ($\sim 10^6$A/m$^2$)  are ultra small compared to those used to manipulate domain walls in ferromagnets ($\sim 10^{11-12}$A/m$^2$) \cite{Jonietz10spin,Schulz12emergent,Yu12skyrmion,Parkin08magnetic,Yamanouchi04current}. 

The motion of the skyrmion in a two dimensional film can be described by a modified version of Newton's equation. For sufficiently slowly varying and not too strong forces, a symmetry analysis suggests the following form of the equations of motion,
%
%
\begin{equation}\label{eom1}
\bm{\mathcal{G}} \times  \dot{\bR}+\alpha \mathcal D \, \dot \bR + m \ddot {\bR} + \alpha \mathbf \Gamma \times \ddot{\bR}  =\bF_{c}+\bF_{g}+\bF_{\rm th}.
\end{equation}
%
%
\AR{Here we assumed translational and rotational invariance of the linearized equations of motion.} The `gyrocoupling' $\bm{\mathcal{G}} = \mathcal G\,\bm{\hat{e}}_{\perp}$ is an effective magnetic field oriented perpendicular to the plane, $\alpha$ is the (dimensionless) Gilbert damping of a single spin, $\alpha \mathcal D$ describes the friction of the skyrmion, $m$ its mass and $\bm R$ its centre coordinate. $\bm \Gamma$ parametrizes a peculiar type of damping proportional to the acceleration of the particle. We name this term `gyrodamping', since it describes the damping of a particle on a cyclotron orbit (an orbit with $ \ddot{\bR} \propto  \bm{\mathcal{G}} \times  \dot{\bR}$), which can be stronger ($\bm \Gamma$ parallel to $\bm{\mathcal{G}}$) or weaker (antiparallel to $\bm{\mathcal{G}}$) than that for linear motion. Our main goal will be to describe the influence of forces on the skyrmion arising from electric currents ($\bf F_c$), magnetic field gradients $(\bf F_g$) and thermal fluctuations $(\bf F_{\rm th}$).

By analyzing the motion of a {\em rigid} magnetic structure $\bM({\bm r}, t) = \bM_0({\bm r}- {\bm R}(t))$ for {\em static} forces, one can obtain analytic formulas for $\mathcal{G}, \alpha \mathcal D, \bm F_c$  and $\bm F_g$ using the approach of Thiele \cite{Thiele73steady, Everschor11current, Schulz12emergent, Everschor12rotating, Iwasaki13universal}. In Ref. [\onlinecite{Makhfudz12inertia}],  an approximate value for the mass of a skyrmion was obtained by simulating the motion of a skyrmion in a nanodisc and by estimating contributions to the mass from internal excitations of the skyrmion.
\begin{figure}[t]
\centering
\includegraphics[width=0.98 \linewidth]{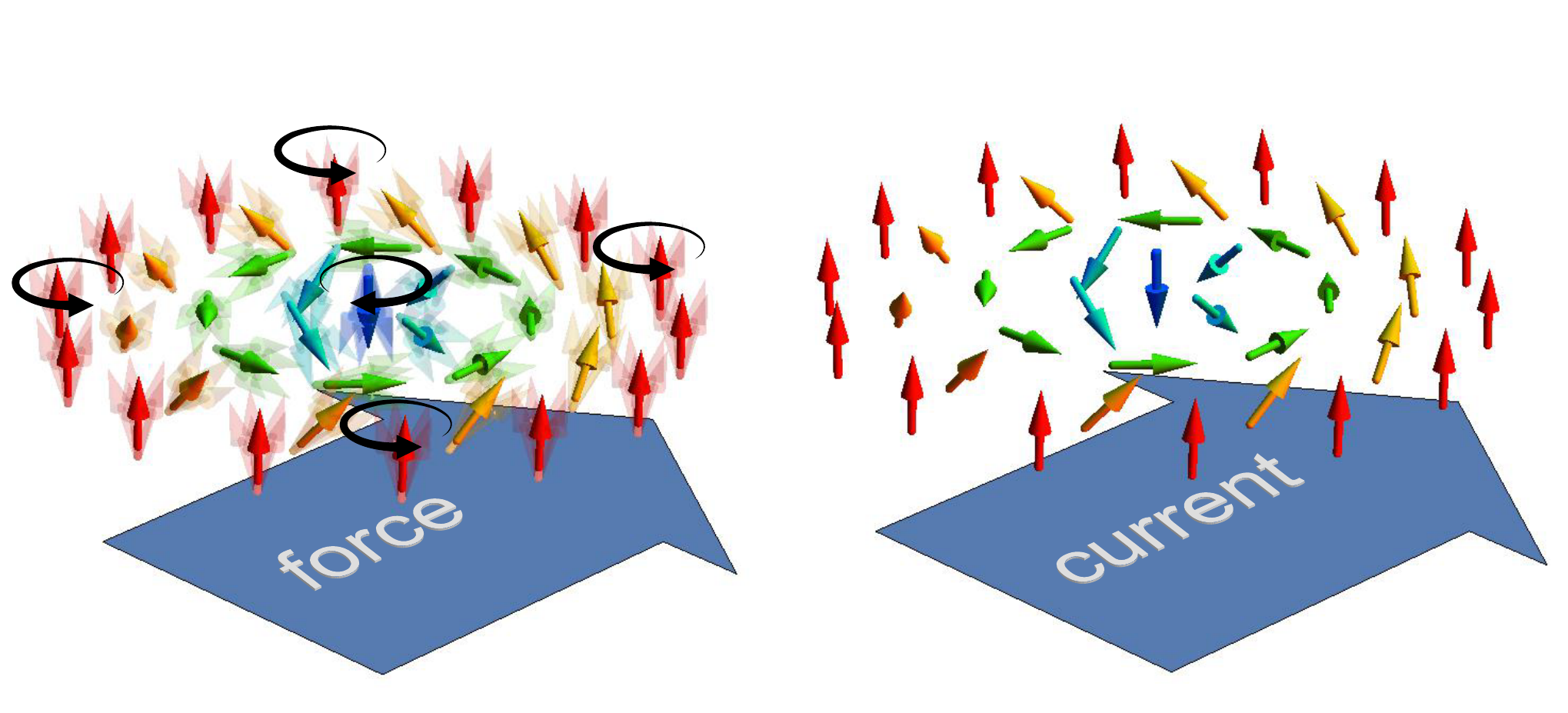}
\caption{\label{figure1}When a skyrmion is driven by a time dependent external force, it becomes distorted and the spins precess resulting in a delayed response and a large effective mass. In contrast, when the skyrmion motion is driven by an electric current, the skyrmion approximately flows with the current with little distortion and precession. Therefore skyrmions respond quickly to rapid changes of the electric current.}
\end{figure}

For rapidly changing forces, needed for the manipulation of skyrmions in spintronic devices, Eq.~(\ref{eom1}) is however not sufficient. A generalized version of Eq. (\ref{eom1}) valid for weak but also arbitrarily time-dependent forces can be written as
%
%
\begin{eqnarray}\label{eom2}
\uu{\bG}^{-1}(\omega) \bV(\omega) &=& \bF_{c}(\omega)+\bF_{g}(\omega)+\bF_{th}(\omega)\\
&=&\uu{S}_c(\omega) \bm  v_s(\omega)+\uu{S}_g(\omega) \bm \nabla B_z(\omega)+{\bF}_{\rm th}(\omega)\nonumber 
\end{eqnarray}
%
%
Here $\bm V(\omega)=\int e^{i \omega t} \dot {\bm R}(t) dt$ is the Fourier transform of the velocity of the skyrmion, $\bm v_s(\omega)$ is the (spin-) drift velocity of the conduction electrons, directly proportional to the current, $\bm \nabla B_z(\omega)$ describes a magnetic field gradient in frequency space. The role of the random thermal forces, ${\bF}_{\rm th}(\omega)$, is special as their dynamics is directly linked via the fluctuation-dissipation theorem to the left-hand side of the equation, see below. The $2 \times 2$ matrix $\bm G^{-1}(\omega)$ describes the dynamics of the skyrmion; its small-$\omega$ expansion defines the terms written on the left-hand side of Eq.~(\ref{eom1}). One can expect strongly frequency-dependent dynamics for the skyrmion because the external forces in combination with the motion of the skyrmion can induce a precession of the spin and also excite spinwaves in the surrounding ferromagnet, see Fig. \ref{figure1}. We will, however, show that the frequency dependence of the right-hand side of the Eq.~(2) is at least as important: not only the motion of the skyrmion but also the external forces excite internal modes. Depending on the frequency range, there is an effective screening or antiscreening of the forces described by the matrices $\uu{S_c}(\omega)$ and $\uu{S_g}(\omega)$. Especially for the current-driven motion, there will be for all frequencies an almost exact cancellation of terms from $\uu{\bG}^{-1}(\omega)$ and  $\uu{S_c}(\omega)$. As a result the skyrmion will follow almost instantaneously any change of the current despite its large mass.

In this paper, we study the dynamics of a driven skyrmion by solving numerically the stochastic Landau-Lifshitz-Gilbert (LLG) equation. \AR{Our strategy will be to determine the parameters of Eq.~(\ref{eom2}) such that this equation reproduces the results of the LLG equation.}  Section \ref{sec:Model} introduces the model and outlines the numerical implementation. Three driving mechanisms are considered: section \ref{sec:ThermalDiffusion} studies the diffusive motion of the skyrmion due to thermal noise, section \ref{sec:ForceDrivenmotion} the skyrmion motion due to time-dependent magnetic field gradient and section \ref{sec:CurrentDrivenMotion} the current-driven dynamics. We conclude with a summary and discussion of the results in Sec. \ref{sec:Conclusion}.

\section{Model} 
\label{sec:Model}
Our study is based on a numerical analysis of the stochastic Landau-Lifshitz-Gilbert (sLLG) equations \cite{Garcia98langevin} defined by
%
%
\begin{eqnarray}\label{eq:sLLG}
\frac{d\bM_\br}{dt} = &\gamma& \bM_\br \times \left[ \bB_{\rm eff} + \bb_{\rm fl}(t) \right]  \nonumber \\ &-& \gamma \frac{\alpha}{M} \bM_\br \times \left( \bM_\br \times \left[ \bB_{\rm eff} + \bb_{\rm fl}(t)  \right] \right).
\end{eqnarray}
%
%
Here $\gamma$ is the gyromagnetic moment and $\alpha$ the Gilbert damping; $\bB_{\rm eff} = -\frac{\delta H[\bM]}{\delta \bM_\br}$ is an effective magnetic field created by the surrounding magnetic moments and $\bb_{\rm fl}(t)$ a fluctuating, stochastic field creating random torques on the magnetic moments to model the effects of thermal fluctuations, see below.

The Hamiltonian $H[\bM]$ is given by
%
%
\begin{eqnarray}\label{Hamiltonian}
H[\bM] &=& -J \sum_{\br} \bM_\br \cdot \left( \bM_{\br+a\be_x} + \bM_{\br+a\be_y} \right)\nonumber\\
&-& \lambda\sum_{\br} \left( \bM_\br \times \bM_{\br+a\be_x} \cdot \be_x + \bM_\br \times \bM_{\br+a\be_y} \cdot \be_y \right)\nonumber\\
&-& \bB \cdot \sum_\br \bM_\br
\end{eqnarray}
%
%
We use $J=1$, $\gamma=1$, $|\bM_\br|=1$, $\lambda = 0.18 J$ for the strength of the Dzyaloshinskii-Moriya interaction and $\bB=(0,0,0.0278 J)$ for all plots giving rise to a skyrmion with a radius of about $15$ lattice sites, see Appendix \ref{sec:SkyrmionCentre}.
For this parameter set, the ground state is ferromagnetic, thus the single skyrmion is a topologically protected, metastable excitation.
Typically we simulate $100 \times 100$ spins for the analysis of diffusive and current driven motion and $200 \times 200$ spins for the force-driven motion. For these parameters lattice effects are negligible, see appendix \ref{sec:Scaling}. Typical microscopic parameters used, are $J=1\text{ meV}$ (this yields $T_c \sim 10\text{ K}$) which we use to estimate typical time scales for the skyrmion motion.

Following Ref.~\onlinecite{Garcia98langevin}, we assume that the field $\bb^\text{fl}_\br(t)$ is generated from a Gaussian stochastic process with the following statistical properties
%
%
\begin{eqnarray}\label{eq:bfl_stat}
&\phantom{a}&\left\langle \bb^\text{fl}_{\br,i}(t) \right\rangle = 0 \nonumber \\
&\phantom{a}&\left\langle \bb^\text{fl}_{\br,i}(t) \bb^\text{fl}_{\br',j}(s) \right\rangle = 2 \alpha \frac{k_B T}{\gamma M} \delta_{ij} \delta_{\br \br'} \delta(t-s)
\end{eqnarray}
%
%
where $i$ and $j$ are cartesian components and $\langle\dots\rangle$ denotes an average taken over different realizations of the fluctuating field. The Gaussian property of the process stems from the interaction of $\bM_\br$ with a large number of microscopic degrees of freedom (central limit theorem) which are also responsible for the damping described by $\alpha$, reflecting the fluctuation-dissipation theorem. The delta-correlation in time and space in  Eq.~(\ref{eq:bfl_stat}) expresses that the autocorrelation time and length of $\bb^\text{fl}_\br(t)$ is much shorter than the response time and length scale of the magnetic system. 

For a numerical implementation of Eq.~(\ref{eq:sLLG}) we follow Ref.~\onlinecite{Garcia98langevin} and use Heun's scheme for the numerical integration which converges quadratically to the solution of the general system of stochastic differential equations (when interpreted in terms of the Stratonovich calculus).

For static driving forces, one can calculate the drift velocity $\dot{\bR}$ following Thiele \cite{Thiele73steady}. Starting from the Landau-Lifshitz Gilbert equations, Eq.~(\ref{eq:sLLG}), we project onto the translational mode by multiplying Eq.~(\ref{eq:sLLG}) with $\partial_i \bM_\br$ and integrating over space \cite{Everschor11current, Everschor12rotating, He2006current}. 
%
%
\begin{eqnarray}\label{eqn:GD}
\mathcal{G} &=& \hbar M_0\, 
 \int d {\bm r}\,  {\bm n} \cdot \left(  { \partial_x {\bm n} }  \times   { \partial_y {\bm n} } \right) \nonumber
\\{\cal D} &=& \hbar M_0 \int d {\bm r} \, ({ \partial_x {\bm n} } \cdot   { \partial_x {\bm n} }+{ \partial_y {\bm n} } \cdot   { \partial_y {\bm n} })/2 \nonumber \\
\bm F_c&=&\bm{\mathcal{G}} \times \bm v_s+ \beta \mathcal D \, \bm v_s , \nonumber \\
\bm F_g&=& M_{\rm s} \bm \nabla B,  \qquad  M_s=M_0\,\int d \bm r \, (1-n_z)
\end{eqnarray}
%
%
where $\bm n$ is the direction of the magnetization, $M_0$ the local \AR{spin density}, $\bm v_s$ the (spin-) drift velocity of the conduction electrons proportional to the electric current, and $M_s$ is the change of the magnetization induced by a skyrmion in a ferromagnetic background. The 'gyrocoupling vector' $\bm{\mathcal{G}} = (0,0,\mathcal{G})^{\rm T}$ with $\mathcal{G}=\pm \hbar M_0 4 \pi$ is given by the winding number of the skyrmion, independent of microscopic details.

\section{Thermal diffusion} 
\label{sec:ThermalDiffusion}
Random forces arising from thermal fluctuations play a decisive role in controlling the diffusion of particles and therefore also the trajectories $\bm R(t)$ of a skyrmion. To obtain $\bm R(t)$ and  corresponding correlation functions we used numerical simulations based on the stochastic Landau-Lifshitz-Gilbert equation~\cite{Garcia98langevin}. These micromagnetic equations describe the dynamics of coupled spins  including the effects of damping and thermal fluctuations. Initially, a skyrmion spin-texture is embedded in a ferromagnetic background. By monitoring the change of the magnetization, we track the center of the skyrmion $\bm R(t)$, see appendix \ref{sec:SkyrmionCentre} for details.

\AR{Our goal is to use this data to determine the matrix $\bm G^{-1}(\omega)$ and the randomly fluctuating thermal forces, $\bm F_{\rm th}(\omega)$, which together fix the equation of motion, Eq.  (\ref{eom2}), in the presence of thermal fluctuations  ($\nabla B_z=v_s=0$). One might worry that this problem does not have a unique solution as both the left-hand and the right-hand side of Eq. (\ref{eom2}) are not known a priori. Here one can, however, make use of the fact that Kubo's fluctuation-dissipation theorem \cite{Kubo66fluctuation} constraints  the thermal forces on the skyrmion described by ${\bF}_{\rm th}$ in  Eq. (\ref{eom2}) by linking them directly to the dissipative contributions of $\bm G^{-1}$. On average $\langle \bF_{\rm th}=0 \rangle$, but its autocorrelation is proportional to the temperature and friction coefficients. In general it is given by 
\begin{align}
\langle \bF_{\rm th}^i(\omega)  \bF_{\rm th}^j(\omega') \rangle = k_B T [\bm G^{-1}_{ij}(\omega)+\bm G^{-1}_{ji}(-\omega)] 2 \pi \delta(\omega+\omega').\label{fluct}
\end{align}
For small $\omega$ one obtains $\langle \bF_{\rm th}^x(\omega)  \bF_{\rm th}^x(\omega') \rangle= 4 \pi k_B T\,  \alpha \mathcal D\, \delta(\omega+\omega')$ while off-diagonal correlations arise from the gyrodamping  $\langle \bF_{\rm th}^x(\omega)  \bF_{\rm th}^y(\omega') \rangle= 4 \pi i \omega k_B T \, \alpha \Gamma \, \delta(\omega+\omega')$.
Using Eq.~(\ref{fluct}) and demanding furthermore that the solution of  Eq.~(\ref{eom2}) reproduces the correlation function $\langle  {\dot R}_i(t) {\dot R_j}(t') \rangle$ (or, equivalently, $\langle  ({R}_i(t)- {R_j}(t'))^2 \rangle$) obtained from the micromagnetic simulations, leads to the condition~\cite{Kubo66fluctuation}
%
%
\begin{align}\label{Gij1}
\bm G_{ij}(\omega)=\frac{1}{k_B T} \int_0^\infty \Theta(t-t') &\langle  {\dot R}_i(t) {\dot R_j}(t') \rangle \\ &e^{i \omega (t-t')} d(t-t')  \nonumber.
\end{align}
%
%
We therefore determine first in the presence of thermal fluctuations ($\nabla B_z=v_s=0$) from
 simulations of the stochastic LLG equation (\ref{eq:sLLG}) the correlation functions of the velocities and use those to determine  $G_{ij}(\omega)$ using Eq.~(\ref{Gij1}). After a simple matrix inversion, this fixes the left-hand side of the equation of motion, Eq. (\ref{eom2}), and therefore contains all information on the (frequency-dependent) effective mass, gyrocoupling, damping and gyrodamping of the skyrmion. Furthermore, the corresponding spectrum of thermal fluctuations is given by  Eq.~(\ref{fluct}).}
\begin{figure}[t]
\centering
\includegraphics[width=0.98 \linewidth]{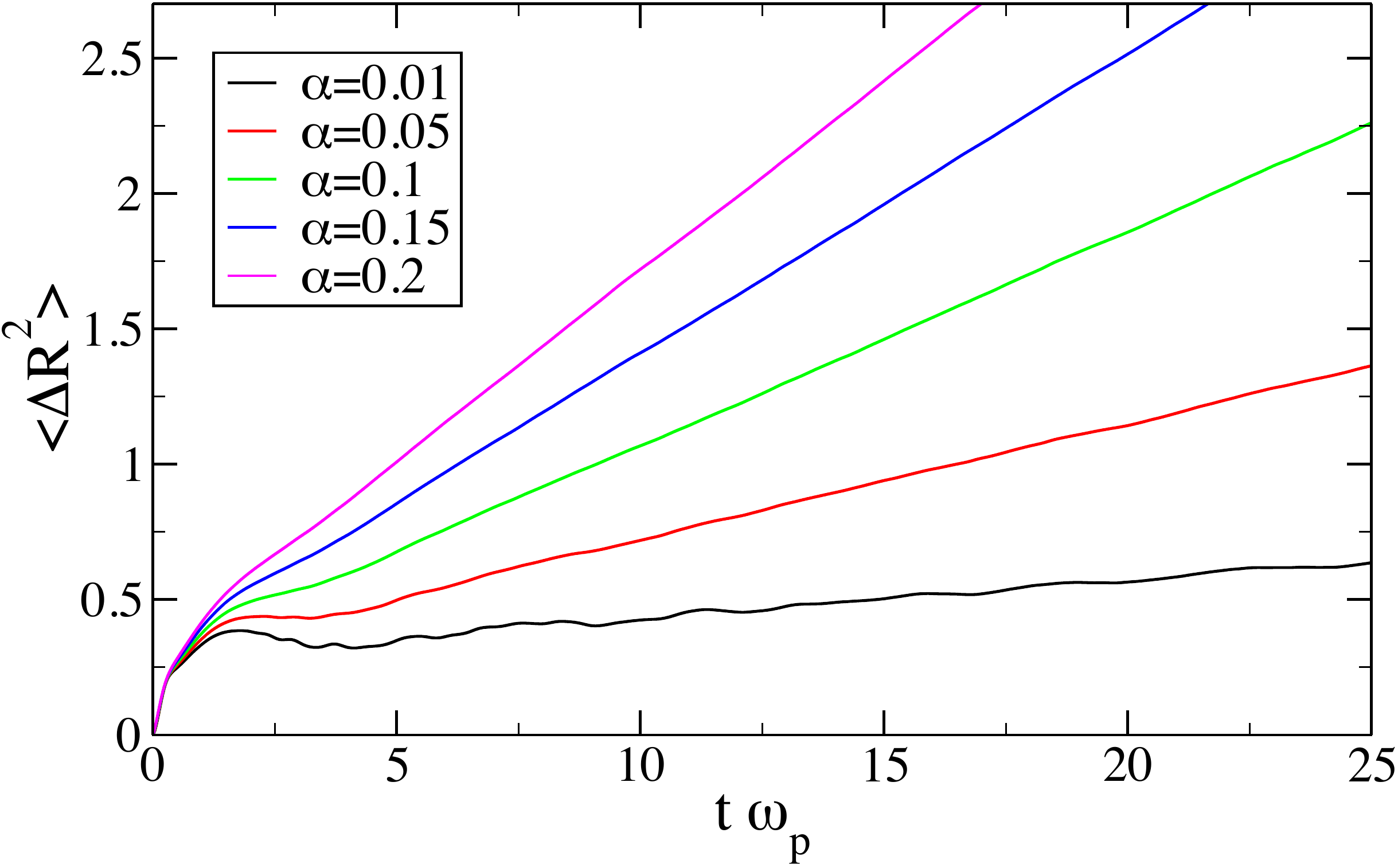}
\caption{\label{figure2}\AR{Time dependence of the correlation function $\left\langle \left(\bR_i(t_0+t) - \bR_i(t_0)\right)^2 \right\rangle$ for $T=0.1 J$ and different values of the Gilbert damping $\alpha$ ($\omega_p=B=0.0278 J$ is the frequency for cyclotron motion). }}
\end{figure}

Fig.~\ref{figure2} shows $\langle (\Delta R)^2 \rangle_t=\langle (R_x(t_0+t)-R_x(t_0))^2 \rangle$. As expected, the motion of the skyrmion is diffusive: the mean squared displacement  grows for long times linearly in time $\langle (\Delta R)^2 \rangle_t= 2 D \, t$, where $D$ is the diffusion constant. Usually the diffusion constant of a particle grows when the friction is lowered~\cite{Kubo66fluctuation}. For the skyrmion the situation is opposite:  the diffusion constant becomes small for the small friction, i.e., small Gilbert damping $\alpha$. This surprising observation has its origin in the gyrocoupling $\mathcal{G}$: in the absence of friction the skyrmion would be localized on a cyclotron orbit. From  Eq.~(\ref{eom1}), we obtain
%
%
\begin{equation}
D=k_B T \frac{\alpha \mathcal D}{\mathcal{G}^2+(\alpha \mathcal D)^2}
\end{equation}
%
%
The diffusion is strongly suppressed by $\mathcal{G}$. As in most materials $\alpha$ is much smaller than unity while $\mathcal{D}\sim \mathcal{G}$, the skyrmion motion is characterized both by a small diffusion constant and a small friction. Such a suppressed dynamics has also been shown to be important for the dynamics of magnetic vortices \cite{clarke}. \AR{For typical parameters relevant for materials like MnSi we estimate that it takes $10^{-6}\,s$ to $10^{-5}\,s$ for a skyrmion to diffusive over an average length of one skyrmion diameter.}
\begin{figure}[t]
\centering
\includegraphics[width=0.98 \linewidth]{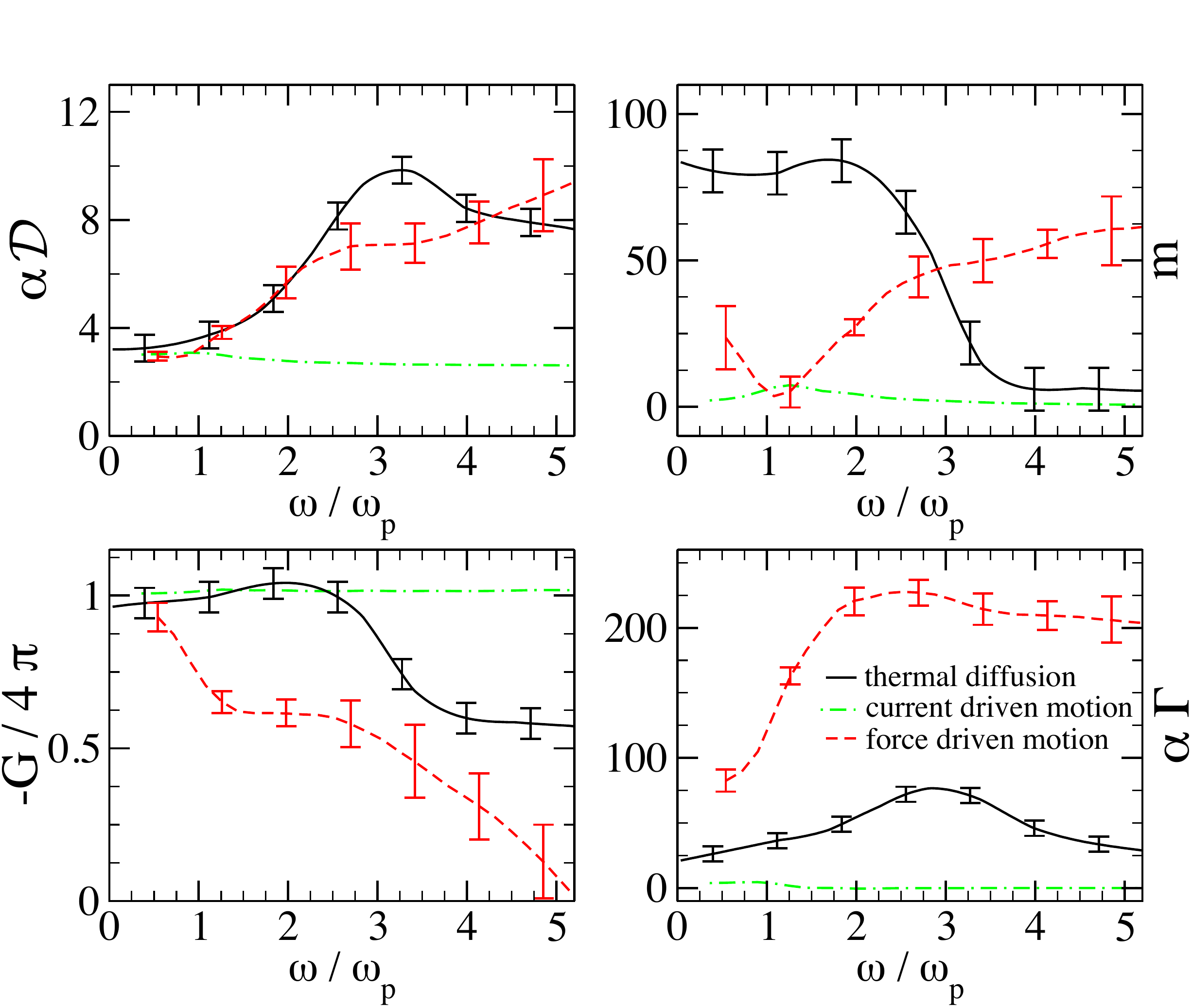}
\caption{\label{figure3}Dissipative tensor $\alpha \mathcal D$, mass $m$, gyrocoupling $\mathcal{G}$ and gyrodamping $\alpha \Gamma$ as functions of the frequency $\omega$ for the diffusive motion at $T=0.1$ (solid lines). They differ strongly from the ``apparent'' dynamical coefficients (see text) obtained for the force driven (red dashed line) and current driven motion (green dot-dashed line). We use $\alpha=0.2$, $\beta=0.1$. The error bars reflect estimates of systematic errors arising mainly from discretization effects, see appendix \ref{sec:Scaling}.}
\end{figure}

To analyze the dynamics on shorter time scales we show in Fig.~\ref{figure3} four real functions parametrizing $\bG^{-1}(\omega)$: a frequency-dependent mass $m(\omega)$, gyrocoupling $\mathcal{G}(\omega)$, gyrodamping $\alpha \Gamma(\omega)$  and dissipation strength $\alpha \mathcal D(\omega)$ with
%
%
\begin{eqnarray}\label{eqn:Gminus1}
\uu{\bG}^{-1}(\omega) = \begin{pmatrix}
\alpha {\cal D}(\omega) - \complex  \omega\, m(\omega) & -\mathcal{G}(\omega) + i \alpha \omega \Gamma(\omega)  \\
 \mathcal{G}(\omega) - i \omega \alpha \Gamma(\omega)  & \alpha {\cal D}(\omega) - \complex \omega \, m(\omega) \nonumber
\end{pmatrix}
\end{eqnarray}
\begin{figure}[t]
\centering
\includegraphics[width=0.98 \linewidth]{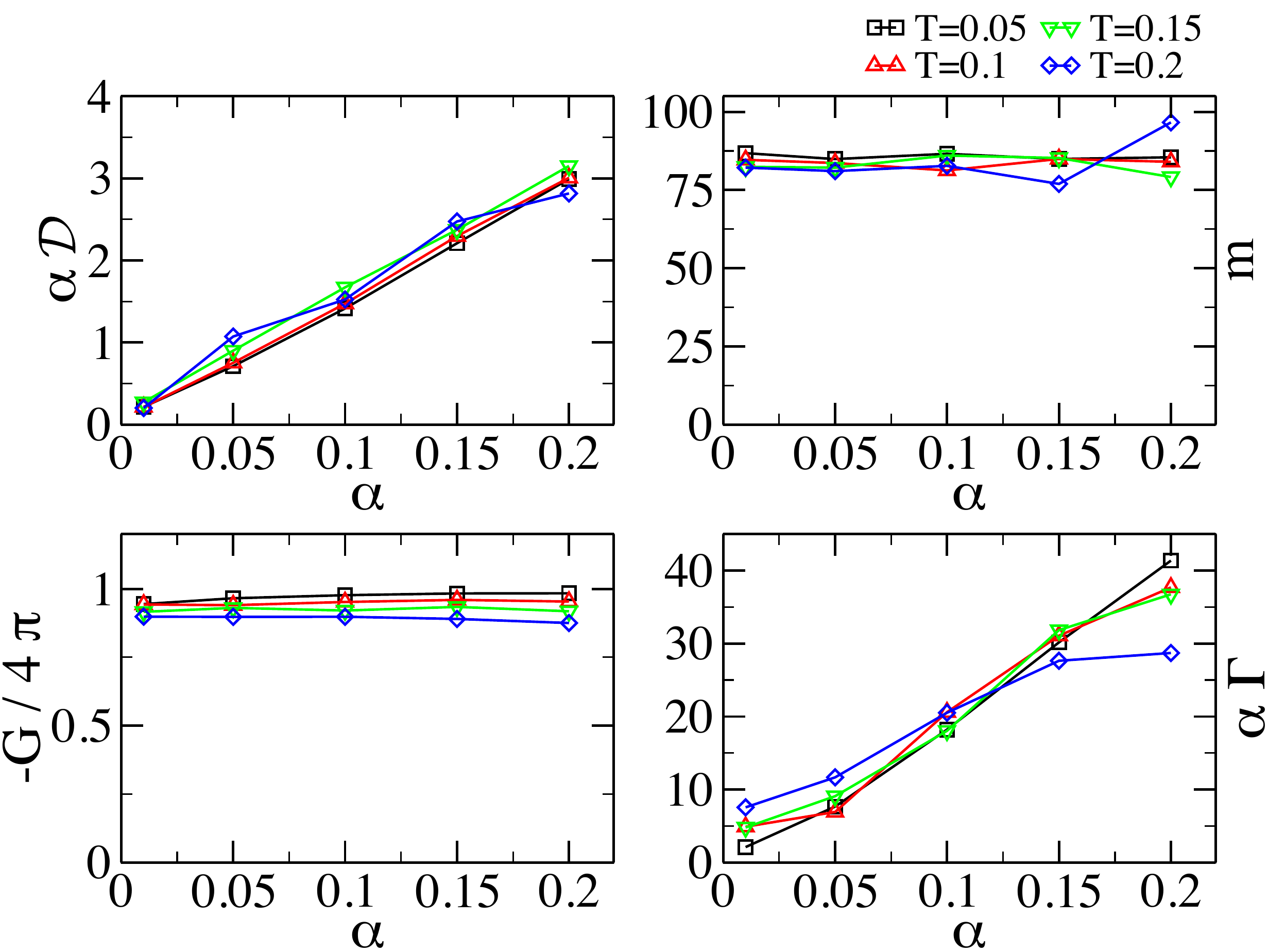}
\caption{\label{figure4}Dissipative strength $\alpha \mathcal D$, mass $m$, gyrocoupling $\mathcal{G}$ and gyrodamping $\alpha \Gamma$ as functions of the Gilbert damping $\alpha$ for different temperatures $T$.}
\end{figure}

For $\omega \to 0$ one obtains the parameters of Eq.~(\ref{eom1}). All parameters depend only weakly on temperature, $\mathcal{G}$ and $m$ are approximately independent of $\alpha$, while the friction coefficients $\alpha \mathcal D$ and $\alpha \Gamma$ are linear in $\alpha$, see Fig.~\ref{figure4}. In the limit $T\to 0$, $\mathcal{G}(\omega \to 0)$ takes the value $-4 \pi$, fixed by the topology of the skyrmion \cite{Thiele73steady,Jonietz10spin}.

Both the gyrodamping $\Gamma$ and and the effective mass $m$ have huge numerical values. A simple scaling analysis of the Landau-Lifshitz-Gilbert equation reveals that both the gyrocoupling $\mathcal{G}$ and $\mathcal{D}$ are independent of the size of the skyrmion, while $\Gamma$ and $m$ are proportional to the area of the skyrmion, and frequencies scale with the inverse area, see appendix \ref{sec:Scaling}. For the chosen parameters (the field dependence is discussed in the appendix \ref{sec:Scaling}), we find $m \approx 0.3\, N_{\rm flip}\, m_0$ and $\alpha\Gamma\approx \alpha \, 0.7 N_{\rm flip} \, m_0$, where $m_0=\frac{\hbar^2}{J a^2}$ is the mass of a single flipped spin in a ferromagnet ($1$ in our units) and we have estimated the number of flipped spins, $N_{\rm flip}$, from the total magnetization of the skyrmion relative to the ferromagnetic background.  As expected the mass of skyrmions grows with the area (consistent with an estimate\cite{petrova} for $m$ obtained from the magnon spectrum of skyrmion crystals), the observation that the damping rate $\alpha \mathcal{D}$ is independent of the size of skyrmions is counter-intuitive. The reason is that larger skyrmions have smoother magnetic configurations, which give rise to less damping. For realistic system parameters $J=1\text{ meV}$ (which yields a paramagnetic transition temperature $T_C  \sim 10$ K, but there are also materials, i.e. FeGe, where the skyrmion lattice phase is stabilised near room-temperature \cite{Yu12skyrmion}) and $a=5\text{ \AA}$ and a skyrmion radius of $~ 200\text{ \AA}$ one finds a typical mass scale of $10^{-25}\text{ kg}$.

The sign of the gyrodamping $\alpha \Gamma$ is opposite to that of the gyrocoupling $\mathcal{G}$. This implies that  $\alpha \Gamma$ describes not damping but rather \textit{anti}damping: there is less friction for cyclotron motion of the skyrmion than for the linear motion. The numerical value for the antidamping turns out to be so large that $\mathcal D m+\Gamma \mathcal{G}<0$. This has the profound consequence that the simplified equation of motion shown in Eq.~(\ref{eom1}) cannot be used: it would wrongly predict that some oscillations of the skyrmion are not damped, but grow exponentially in time due to the strong antidamping. This is, however, a pure artifact of ignoring the frequency dependence of $\bm G^{-1}(\omega)$, and such oscillations do not grow. 

Fig.~\ref{figure3} shows that the dynamics of the skyrmion has a strong frequency dependence. We identify the origin of this frequency dependence with a coupling of the skyrmion coordinate to pairs of magnon excitations as discussed in Ref.~\onlinecite{Schuette14scattering}. Magnon emission sets in for $\omega >2 \omega_p$ where $\omega_p=B$ is the precession frequency of spins in the ferromagnet  (in the presence of a bound state with frequency $\omega_b$, the onset frequency is $\omega_p+\omega_b$, Ref.~\onlinecite{Schuette14scattering}). This new damping channel is most efficient when the emitted spin waves have a wavelength of the order of the skyrmion radius.
\begin{figure}[t]
\centering
\includegraphics[width=0.98 \linewidth]{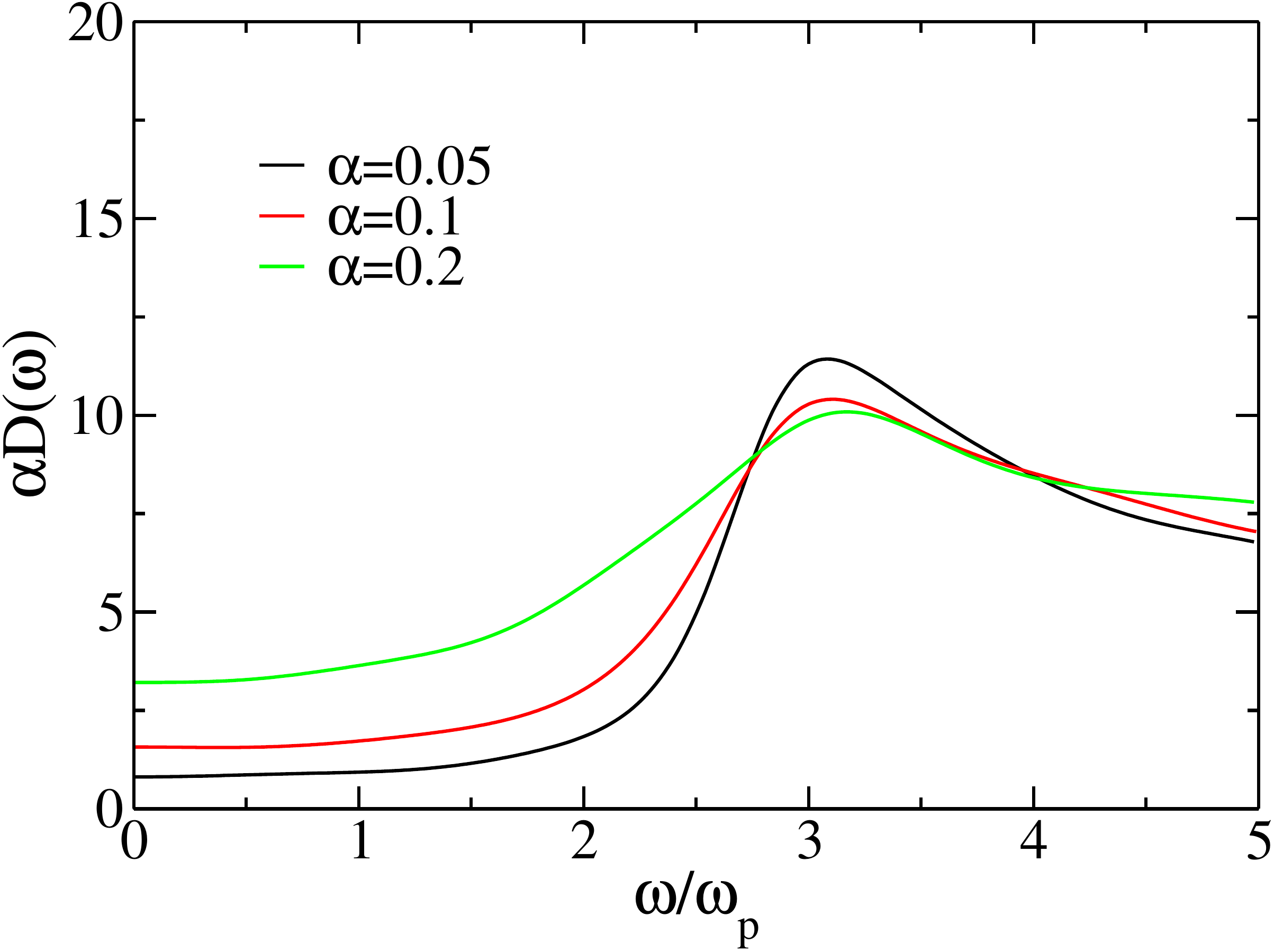}
\caption{\label{figure5} Effective damping, $\alpha \mathcal D(\omega)$ for $\alpha=0.2$, $0.1$ and $0.05$.}
\end{figure}

As a test for this mechanism, we have checked that only this high-frequency damping survives for $\alpha \to 0$. In Fig. \ref{figure5} we show the frequency dependent damping  $\alpha \mathcal D(\omega)$ for various bare damping coefficients $\alpha$. For small $\omega$ it is proportional to $\alpha$ as predicted by the Thiele equation. For $\omega>2 \omega_p$, however, an extra dampling mechanism sets in: the skyrmion motion can be damped by the emission of pairs of spin waves. This mechanism is approximately independent of $\alpha$ and survives in the $\alpha \to 0$ limit. This leads necessarily to a pronounced frequency dependence of the damping and therefore to the effective mass $m(\omega)$ which is related by the Kramers-Kronig relation $m(\omega)=\frac{1}{\omega} \int_{-\infty}^{\infty} \frac{\alpha \mathcal D(\omega')}{\omega'-\omega} \frac{d \omega'}{\pi}$ to $\alpha \mathcal D(\omega)$. Note also that the large $\alpha$ independent mass $m(\omega \to 0)$ is directly related to the $\alpha$ independent damping mechanism for large $\omega$. Also the frequency dependence of $m(\omega)$ and $G(\omega)$ can be traced back to the same mechanism as these quantities can be computed from $\alpha \mathcal D(\omega)$ and $\alpha \Gamma(\omega)$ using Kramers-Kronig relations. For large frequencies, the effective mass practically vanishes and the `gyrocoupling' $\mathcal{G}$ drops by a factor of a half.   

\section{Force-driven motion} 
\label{sec:ForceDrivenmotion}
Next, we study the effects of an oscillating magnetic field gradient $\bm \nabla B_z(t)$ in the absence of thermal fluctuations. As the skyrmion has a large magnetic moment $M^z_{\rm tot}$ relative to the ferromagnetic background, the field gradient leads to a force acting on the skyrmion. In the static limit, the force is exactly given by 
%
%
\begin{equation}
\bm F_g(\omega \to 0)=M^z_{\rm tot} \bm \nabla B_z. 
\end{equation}
\begin{figure}[t]
\centering
\includegraphics[width=0.98 \linewidth]{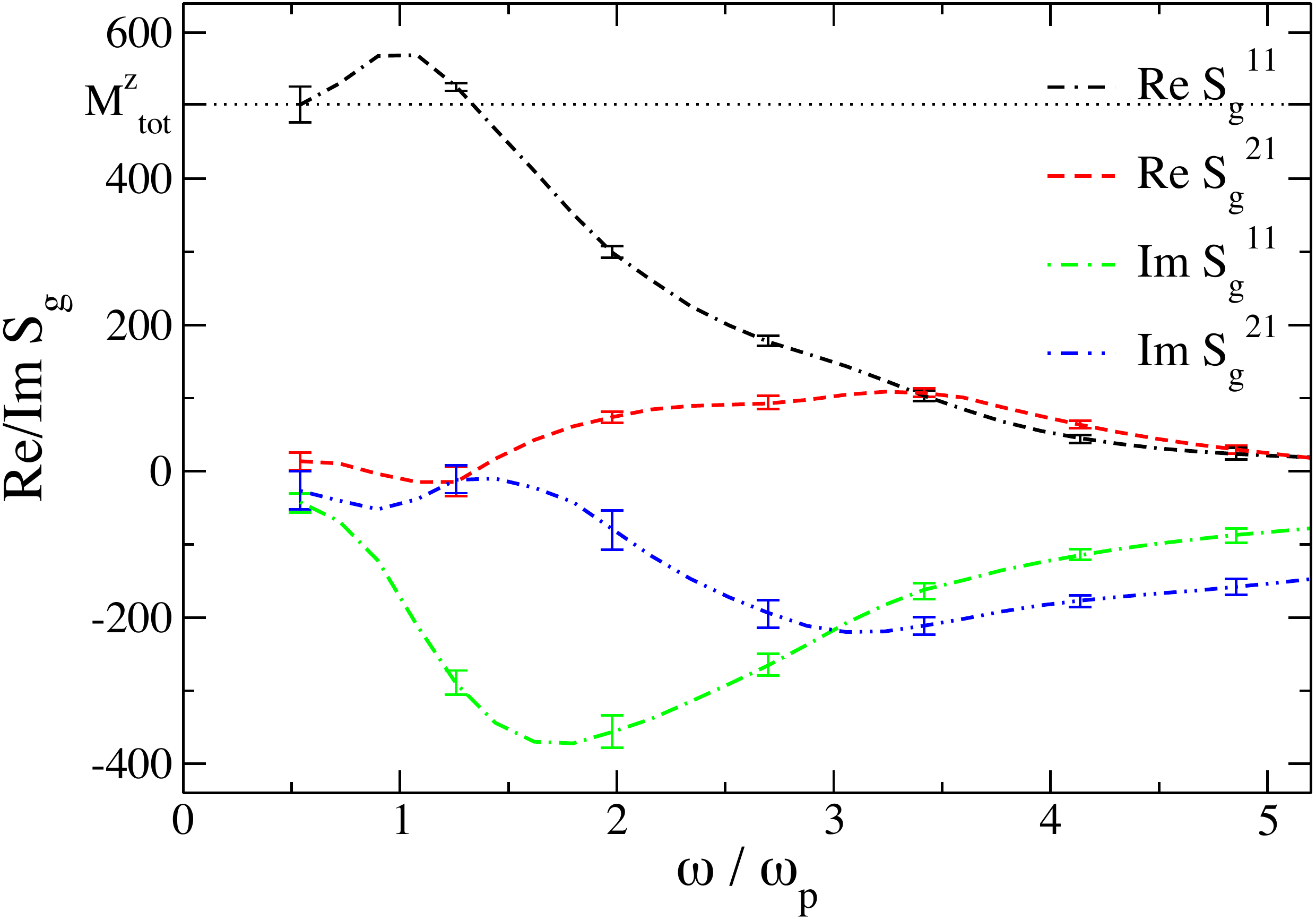}
\caption{\label{figure6}Dynamical coupling coefficients for the force driven motion ($\alpha=0.2$). In the static limit everything but the real part of the diagonal vanishes. ${\text Re}\,S_g^{11}(\omega)$ however approaches the total magnetization $M_{\rm tot}^z$ as expected. The error bars reflect estimates of systematic errors, see appendix \ref{sec:Scaling}.}
\end{figure}
Using $\bG^{-1}(\omega)$ determined above, we can calculate how the effective force $\bm S_g(\omega) \bm \nabla B_z(\omega)$ (see Eq.~\ref{eom2}) depends on frequency. Fig.~\ref{figure6} shows that for $\omega \to 0$ one obtains the expected result $S_g(\omega\to 0)=\delta_{ij} M^z_{\rm tot}$, while a strong frequency dependence sets in above the magnon gap, for $\omega \gtrsim \omega_p$. This is the precession frequency of spins in the external magnetic field.

In general, both the screening of forces (parametrized by $\bf S_g(\omega)$) and the internal dynamics (described by $\bf G^{-1}(\omega)$) determines the response of skyrmions, $\bm V(\omega)=\bm G(\omega)  \bm S_g(\omega) \bm \nabla B_z(\omega)$. Therefore it is in general not possible to extract, e.g., the mass of the skyrmion as described by $\bm G^{-1}(\omega)$ from just a measurement of the response to field gradients. It is, however, instructive to ask what ``apparent'' mass one obtains, when the frequency dependence of $\bf S_g(\omega)$ is ignored. We therefore define the ``apparent'' dynamics $\bm G_a^{-1}(\omega)$ by $\bm G_a(\omega)  \bm S_g(\omega=0)=\bm G(\omega)  \bm S_g(\omega)$. The matrix elements of $\bm G_a^{-1}(\omega)$ are shown in Fig.~\ref{figure3} as dashed lines. The apparent mass for gradient-driven motion, for example, turns out to be more than a factor of three smaller then the value obtained from the diffusive motion clearly showing the importance of screening effects on external forces. The situation is even more dramatic when skyrmions are driven by electric currents.

\section{Current-driven motion} 
\label{sec:CurrentDrivenMotion}
Currents affect the motion of spins both via adiabatic and non-adiabatic spin torques\cite{Tatara08microscopic}. Therefore one obtains two types of forces on the spin texture even in the static limit \cite{Thiele73steady, Everschor11current, Schulz12emergent, Everschor12rotating, Iwasaki13universal}.

The effect of a time-dependent, spin-polarized current on the magnetic texture can be modelled by supplementing the right hand side of eq. (\ref{eq:sLLG}) with a spin torque term $\bT_{\text{ST}}$,
%
%
\begin{eqnarray}\label{eq:cd_LLG}
\bT_{\text{ST}} = -({\bm v}_s\cdot\nabla) \bM_\br + \frac{\beta}{M} \left[ \bM_\br \times ({\bm v}_s \cdot \nabla) \bM_\br \right]. 
\end{eqnarray}
%
%
The first term is called the spin-transfer-torque term and is derived under the assumption of adiabaticity: the conduction-electrons adjust their spin orientation as they traverse the magnetic sample such that it points parallel to the local magnetic moment $\bM_\br$ owing to $J_H$ and $J_{sd}$. This assumptions is justified as the skyrmions are rather large smooth objects (due to the weakness of spin-orbit coupling). The second so called $\beta$-term describes the dissipative coupling between conduction-electrons and magnetic moments due to non-adiabatic effects. Both $\alpha$ and $\beta$ are small dimensionless constants in typical materials. From the Thiele approach one obtains the force 
%
%
\begin{equation}
\bm F_c(\omega \to 0)=\bm{\mathcal{G}} \times \bm v_s + \beta \mathcal D \, \bm v_s.
\end{equation}
%
%

For a Galilei-invariant system one obtains $\alpha=\beta$. In this special limit, one can easily show that an exact solution of the LLG equations in the presence of a time-dependent current, described by $\bm v_s(t)$ is given by $\bm M(\bm r - \int_{-\infty}^t \bm v_s(t') dt')$ provided, $\bm M(\vec r)$ is a static solution of the LLG equation for $\bm v_s= \bm 0$. This implies that for $\alpha=\beta$, the skyrmion motion exactly follows the external current, $\dot{\bm R}(t) = \bm v_s(t)$. Using Eq.~(\ref{eom2}), this implies that for $\alpha=\beta$ one has $\bm G^{-1}(\omega)=\bm S_c(\omega)$. Defining the apparent dynamics, as above, $\bm G_a(\omega) \bm S_c(\omega=0)=\bm G(\omega)  \bm S_c(\omega)$ one obtains a frequency independent $\bm G_a^{-1}(\omega)=S_c(\omega=0)=\beta \mathcal D \id - i \sigma_y \mathcal G$: the apparent effective mass and gyrodamping are exactly zero in this limit and the skyrmion follows the current without any retardation.
\begin{figure}[t]
\centering
\includegraphics[width=0.98 \linewidth]{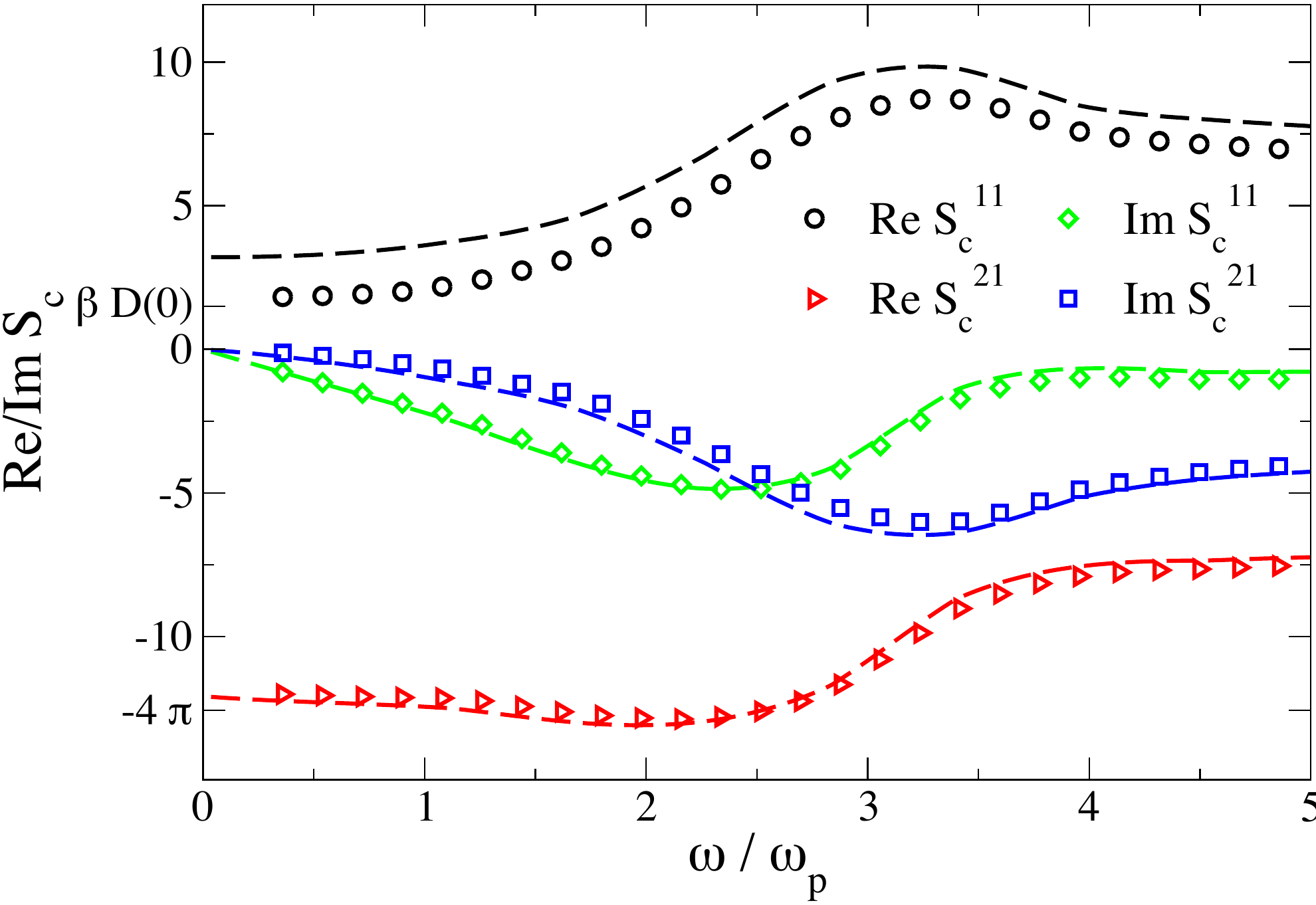}
\caption{\label{figure7}Dynamical coupling coefficients (symbols) for the current-driven motion ($\alpha=0.2$, $\beta=0.1$, $J=1$, $\lambda=0.18 J$, $B=0.0278$). These curves follow almost the corresponding matrix elements of $\bG^{-1}(\omega)$ shown as dashed lines.
A deviation of symbols and dashed line is only sizable for $\text{Re } S^{11}_c$.}
\end{figure}
For $\alpha \neq \beta$, the LLG equations predict a finite apparent mass.  Numerically, we find only very small apparent masses, $m^a_c\propto \alpha-\beta$, see dot-dashed line in upper-right panel of Fig.~\ref{figure3}, where the case $\alpha=0.2$, $\beta=0.1$ is shown. This is anticipated from the analysis of the $\alpha=\beta$ case: As the mass vanishes for $\alpha=\beta=0$, it will be small as long as both  $\alpha$ and $\beta$ are small. Indeed even for $\alpha\neq \beta$ this relation holds approximately as shown in Fig.~\ref{figure7}. The only sizable deviation is observed for $\text{Re } S^{11}_c$ for which the Thiele equation predicts $\text{Re } S^{11}_c(\omega \to 0)=\beta \mathcal D$ while  $\text{Re } {\uu{G}^{-1}}_{11}(\omega \to 0)=\alpha \mathcal D$ as observed numerically.
\begin{figure}[t]
\centering
\includegraphics[width=0.98 \linewidth]{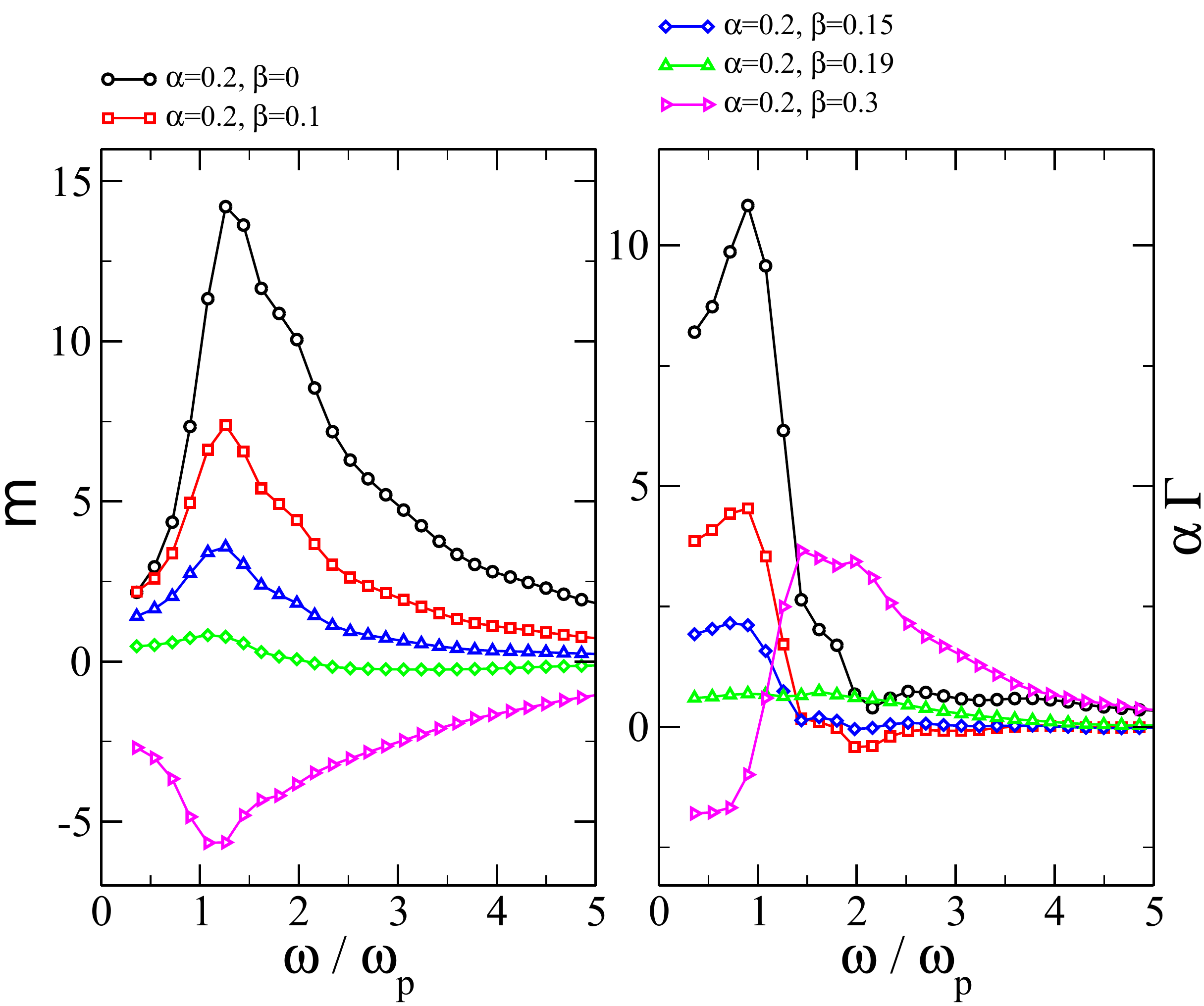}
\caption{\label{figure8}Mass $m(\omega)$ and gyrodamping $\alpha\, \Gamma(\omega)$ as functions of the driving frequency $\omega$ for the current-driven motion. \AR{Note} that both $M$ and $\Gamma$ vanish for $\alpha=\beta$.}
\end{figure}

A better way to quantify that the skyrmion follows the current even for $\alpha\neq \beta$ almost instantaneously is to calculate the apparent mass and gyrodamping for current driven motion, where only results for $\alpha=0.2$ and $\beta=0.1$ have been shown. As these quantities vanish for $\alpha=\beta$, one can expect that they are proportional to $\alpha-\beta$ at least for small $\alpha,\beta$. This is indeed approximately valid at least for small frequencies as can be seen from Fig.~\ref{figure8}. Interestingly, one can even obtain negative values for $\beta>\alpha$ (without violating causality). Most importantly, despite the rather large values for $\alpha$ and $\beta$ used in our analysis, the apparent effective mass and gyrodamping remain small compared to the large values obtained for force-driven motion or the intrinsic dynamics. This shows that retardation effects remain tiny when skyrmions are controlled by currents.

\section{Conclusions} 
\label{sec:Conclusion}
In conclusion, we have shown that skyrmions in chiral magnets are characterised by a number of unique dynamical properties which are not easily found in other systems. First, their damping is small despite the fact that skyrmions are large composite objects. Second, despite the small damping, the diffusion constant remains small. Third, despite a huge inertial mass, skyrmions react almost instantaneously to external currents. The combination of these three features can become the basis for a very precise control of skyrmions by time-dependent currents.

Our analysis of the skyrmion motion is based on a two-dimensional model where only a single magnetic layer was considered. All qualitative results can, however, easily be generalized to a film with $N_L$ layers. In this case, all terms in Eq.~(\ref{eom1}) get approximately multiplied by a factor $N_L$ with the exception of the last term, the random force, which is enhanced only by a factor $\sqrt{N_L}$. As a consequence, the diffusive motion is further suppressed by a factor $1/\sqrt{N_L}$ while the current- and force-driven motion are approximately unaffected.

An unexpected feature of the skyrmion motion is the antidamping arising from the gyrodamping. The presence of antidamping is closely related to another important property of the system: both the dynamics of the skyrmion and the effective forces acting on the skyrmion are strongly frequency dependent.

In general, in any device based on skyrmions a combination of effects will play a role. Thermal fluctuations are always present in room-temperature devices, the shape of the device will exert forces \cite{Iwasaki13constricted, Iwasaki2014colossal} and, finally, we have identified the current as the ideal driving mechanism. In the linear regime, the corresponding forces are additive. The study of non-linear effects and the interaction of several skyrmions will be important for the design of logical elements based on skyrmions and this is left for future works. As in our study, we expect that dynamical screening will be important in this regime.

\begin{acknowledgements}
The authors are greatful for insightful discussions with K. Everschor and Markus Garst. Part of this work was funded through the Institutional Strategy of the University of Cologne
within the German Excellence Initiative" and the BCGS. C.S. thanks the University of Tokyo for hospitality during his research internship where part of this work has been performed.
N.N. was supported by Grant-in-Aids for Scientific Research (No. 24224009) from the Ministry of Education, Culture, Sports, Science and Technology (MEXT) of Japan, and by the Strategic International Cooperative Program (Joint Research Type) from
Japan Science and Technology Agency. J.I. is supported by Grant-in-Aids for JSPS Fellows (No.~2610547).
\end{acknowledgements}

\appendix
\section{Definition of the Skyrmion's centre coordinate}
\label{sec:SkyrmionCentre}
In order to calculate the Green's function, Eq.~(3), one needs to calculate the velocity-velocity correlation function. Therefore it is necessary to track the skyrmion position throughout the simulation. 
Mostly two methods have been used so far for this\cite{Makhfudz12inertia}: (i) tracking the centre of the topological charge and (ii) tracking the core of the Skyrmion (reversal of magnetization).
\begin{figure}[t]
\centering
\includegraphics[width=0.85 \linewidth]{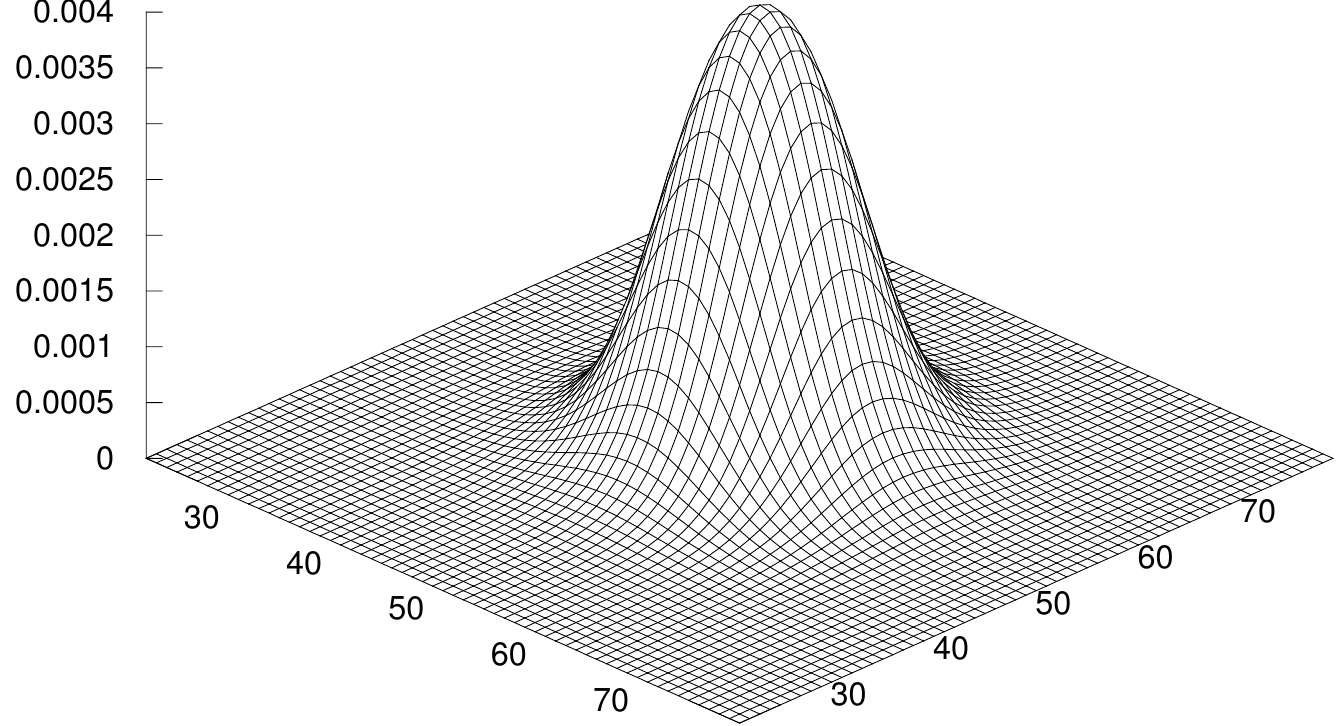}
\caption{\label{fig1Supplement}Skyrmion density based on the normalized $z$-component of the 
magnetization.}
\end{figure}

The topological charge density
%
%
\begin{eqnarray}
\rho_{\rm top}(\mathbf{r}) = \frac{1}{4 \pi} \mathbf{\hat{n}}(\mathbf{r}) \cdot \left(  \partial_x \mathbf{\hat{n}}(\mathbf{r}) \times  \partial_y \mathbf{\hat{n}}(\mathbf{r}) \right)
\end{eqnarray}
%
%
integrates to the number of Skyrmions in the system. Therefore for our case of a single Skyrmion in the ferromagnetic background this quantity is normalized to $1$. The center of topological charge can therefore be defined as
%
%
\begin{eqnarray}
\mathbf{R} = \int d^2\mathbf{r} \rho_{\rm top}(\mathbf{r}) \mathbf{r}
\end{eqnarray}
%
%
For the case of finite temperature this method can, however, not be used directly. Thermal fluctuations in the ferromagnetic background far away from the skyrmion lead to a large noise to this quantity which diverges
in the thermodynamic limit. A similar problem arises when tracking the center using the magnetization of the skyrmion.

One therefore needs a method which focuses only on the region close to the skyrmion center. To locate the skyrmion, we use the $z$-component of the magnetization but take into account only points where  $M_z(\br)<-0.7$ (the magnetization of the ferromagnetic background at $T=0$ is $+1$). We therefore use
%
%
\begin{eqnarray}
\rho(\br) &=& (1-\bM_z(\br)) \, \Theta[-M_z(\br)-0.7]
\end{eqnarray}
%
%
where $\Theta[x]$ is the theta function. 
A first estimate, $\bm R_{\rm est}=R_V$, for the radius is obtained from 
%
%
\begin{eqnarray}
\bm R_{A}=\frac{\int_A \bm r \rho(\br) \,d^2 \bm r}{\int_A \rho(\br)\,d^2 \bm r}
\end{eqnarray}
%
%
by integrating over the full sample volume $V$. $\bm R_{\rm est}$ is noisy due to the problems mentioned above but for the system sizes simulated one nevertheless obtains a good first estimate for the skyrmion position. This estimate is refined by using in a second step for the integration area only $D = \left\{\br \in \mathbb{R}^2\,|\,|\br-\mathbf{R}_{\rm est}|<r \right\}$ where
$r$ is choosen to be larger than the radius of the skyrmion core (we use $r =1.3 \sqrt{N_{<}/\pi}$, where $N_<$ is the number of spins with $M_z<-0.7$). Thus we obtain a reliable estimate, $\bm R=\bm R_D$, not affected by spin fluctuations far away from the skyrmion. From the resulting $\bm R(t)$, one can obtain the velocity-velocity correlation function  $\left\langle \bV_i(t_0+t) \bV_J(t_0) \right\rangle$. 

\section{Scaling invariance}
\label{sec:Scaling}
To obtain analytic insight into the question of how parameters depend on system size and to check the numerics for lattice artifacts it is useful to perform a scaling analysis of the sLLG equations and the effective equations of motion for the skyrmion.

We investigate a scaling transformation, where the radius of the skyrmion is enlarged by a factor $\eta$, $\bM(\br)\to \tilde \bM(\br)=\bM(\br/\eta)$.

For the scaling analysis, we use the continuum limit of Eq.~(4) (setting $a=1$)
%
%
\begin{eqnarray}
H[\bM] = \int d^2\br \left[ \frac{J}{2} \left( \nabla \bM \right)^2 + \lambda \bM \cdot \nabla \times \bM - \bB \cdot \bM \right] \nonumber
\end{eqnarray}
%
%
The three terms scale with $\eta^0$, $\eta$ and $\eta^2$, respectively.  To obtain a larger skyrmion, we therefore have to rescale $\lambda \to \lambda/\eta$ and $\bB\to \bB/\eta^2$. This implies that the $\bm B_{\rm eff}$ term in the sLLG equation scales with $1/\eta^2$ and therefore also the time axis
has to be rescaled, $t \to \eta^2 t$, implying that all time scales are a factor of $\eta^2$ longer and all frequencies a factor $1/\eta^2$ smaller. Similiarly, the driving current, i.e. $\bm v_s$ is reduced by a factor $\eta^{-1}$ while the temperature remains unscaled. This implies that when $\bM(\br,t)$ is a solution for a given value of $\lambda$, $\bB$ and $\bm v_s$ and $\uu{G}(\omega)$ the corresponding velocity-correlation function of the skyrmion, then $\bM(\br/\eta,t/\eta^2)$ is a solution for 
$\lambda/\eta$, $\bB/\eta^2$, $\bm v_s/\eta$ with correlation function $\uu{G}(\omega \eta^2)$.

Accordingly, the $\omega \to 0$ limit and therefore the gyrocoupling $\mathcal G$, the friction constant $\alpha \mathcal D$ and the diffusion constant of the skyrmion are {\em independent} of $\eta$, consistent with the analytical formulas eq.~(6). In contrast, the mass of the skyrmion, $m$, and the gyrodamping $\alpha \Gamma$ scale with $\eta^2$. They are therefore proportional to the number of spins constituting the skyrmion consistent with our numerical findings.
\begin{figure}[t]
\centering
\includegraphics[width=0.85 \linewidth]{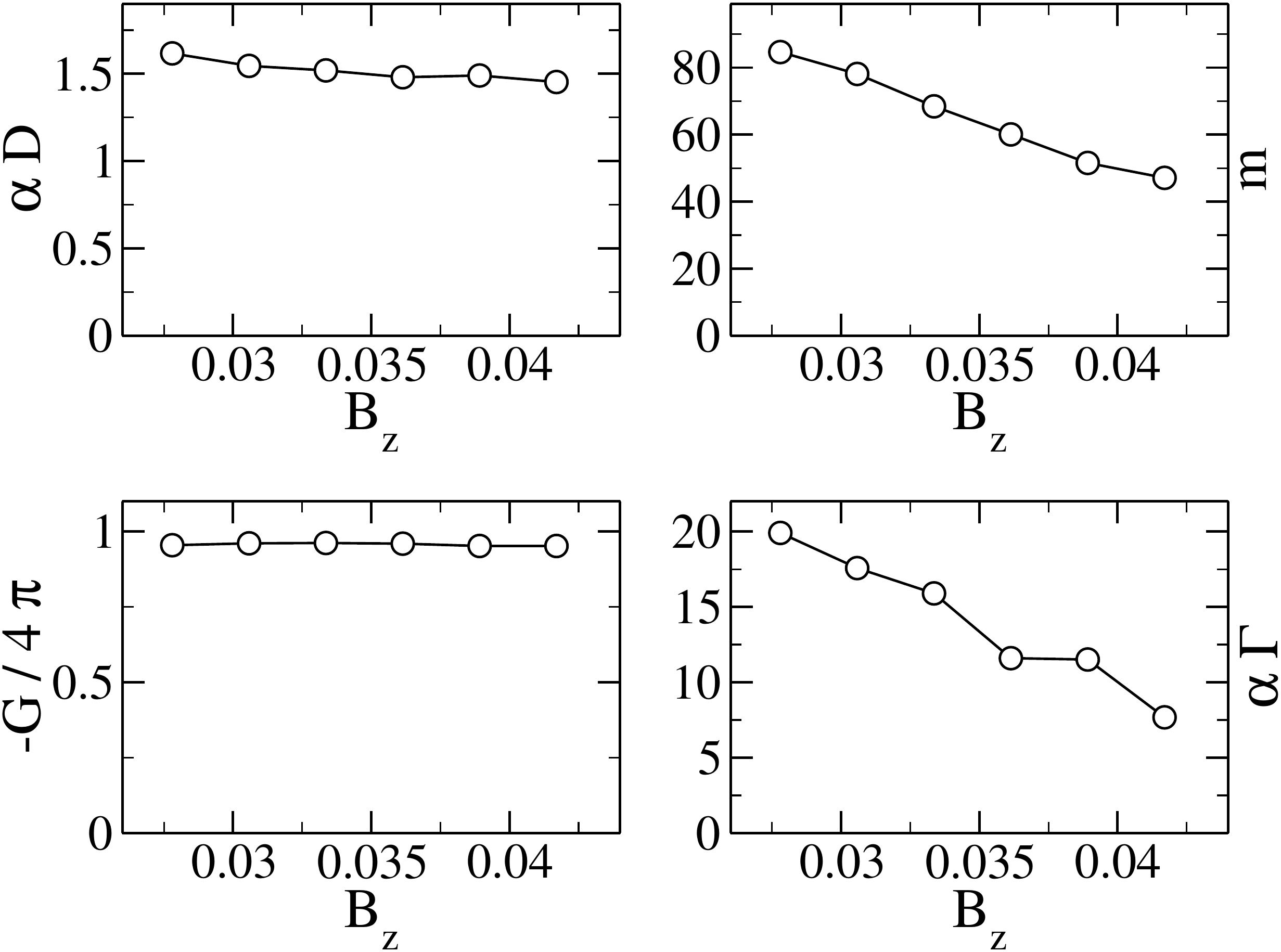}
\caption{\label{fig2Supplement}Field dependence of the dissipative tensor $\alpha \mathcal D$, the mass $m$, the gyrocoupling $\mathcal{G}$ and the gyrodamping $\alpha \Gamma$   for $J=1$, $\lambda=0.18 J$. The main effect is that both $m$ and $\alpha \Gamma$ shrink when the size of the skyrmion shrinks with increasing $B_z$.}
\end{figure}
When one does, however, change the external magnetic field for {\em fixed} strength $\lambda$ of the DM interaction, the internal structure of the skyrmion and therefore also $\uu{G}(\omega)$ change quantitatively in a way not predictable by scaling. In Fig.~\ref{fig2Supplement} we therefore show the $B$-dependence of $\alpha \mathcal D$, $m$, $\mathcal G$ and $\alpha \Gamma$. For increasing $B$ the size of the skyrmion shrinks. From our previous analysis, it is therefore not surprising that both the effective mass $m$ and the gyrodamping $\alpha \Gamma$ shrink substantially while  $\alpha \mathcal D$ and $\mathcal G$ are less affected . Quantitatively, this change of $m$ and $\alpha \Gamma$ is however not proportional to the number of spins constituting the skyrmion. 

We have tested numerically the scaling properties for $\eta=1.2$ and find that all features are quantitatively reproduced. Small variations on the level of a few percent do, however, occur reflecting the typical size of features arising from the discretization of the continuum theory. A conservative estimate of such systematic discretization effects for the diffusive motion is given by the error bars in Figs.~\ref{figure3} (all statistical errors are smaller than the thickness of the line). For the field-driven motion (Fig. \ref{figure3} and Fig. \ref{figure6}) spatial discretization effects lead to a different source of errors. For very small field gradients and high frequencies the displacement of the skyrmion is much smaller than the lattice spacing and the response is affected by a tiny pinning of the skyrmion to the discreet lattice. For larger gradients, however, nonlinear effects set in and for small frequencies the skyrmion starts to approach the edge of the simulated area. In Fig. \ref{figure3}, we therefore used for the force-driven motion $\nabla B=0.0005$ for $\omega< 2 \omega_p$ and $\nabla B=0.0015$ for  $\omega> 2 \omega_p$. Error bars have been estimated from variations of the numerical values when  $\nabla B$ was varied from $0.0001$ to $0.0015$. For the current-driven motion errors are so tiny that they are not shown.


\end{document}